\documentclass{article}
\usepackage{a4,amsmath,amsthm,eucal,eufrak}
\renewcommand{\theequation}{\thesection.\arabic{equation}}
\newcommand{\der}{{\rm d}}
\newcommand{\bom}{\boldsymbol{\omega}}
\newcommand{\mw}{{\scriptstyle\bigwedge}}
\newtheorem*{lem}{Lemma}
\begin{document}
\title{Equations of Geodesic Deviation and the Inverse Scattering
Transform}
\author{V. V. Varlamov\\
{\it\small Department of Mathematics, Siberia State University of Industry},\\
{\it\small Kirova 42, Novokuznetsk 654007, Russia}}
\date{}
\maketitle
\begin{abstract}\begin{sloppypar}
Solutions of equations of geodesic deviation in three- and four-dimensional
spaces are obtained via the inverse scattering transform. It is
shown that in the case of three-dimensional space solutions of geodesic
deviation equations are reduced to solutions of the well-known 
Zakharov-Shabat problem. In four-dimensional space system of geodesic
deviation equations is associated with a $3\times 3$ matrix Schr\"{o}dinger
equation, and dependence on parameters is defined by the nonlinear equations
of three-wave interaction.\end{sloppypar}
\end{abstract}

\section{Introduction}
\setcounter{equation}{0}
It is well-known that a $m\times m$ matrix Schr\"{o}dinger equation on
$-\infty<x<\infty$ is defined by the following expression \cite{1} :
$$L\psi(x,k)=\lambda\psi(x,k),\quad\lambda=k^2,$$
where
\begin{eqnarray}
&&L=-(\partial^2/\partial x^2)I+U(x),\nonumber\\
&&I=(\delta_{ij}),\quad U(x)=\left(u_{ij}(x)\right);\;\;i,j=1,\ldots,m,
\nonumber\\
&&\psi(x,k)=\left[\psi_1(x,k),\psi_2(x,k),\ldots,\psi_m(x,k)\right].\nonumber
\end{eqnarray}

Further, let $\eta^i$ be the components of deviation vector between two
infinitesimally nearby geodesic lines. Then the components $\eta^i$ satisfy
to the Jacoby equation \cite{2}
\begin{equation}\label{i1}
v^i\nabla_i(v^j\nabla_j\eta^l)=-v^iR^l_{ikm}v^m\eta^k,
\end{equation}
where $v^i$ are the components of the tangent vector along a geodesic line
$\gamma$, $R^i_{jkl}$ is the curvature tensor of the metrics
$$ds^2=g_{ij}dx^idx^j.$$
In a special system of coordinates, where axis $x^j$ is a geodesic line,
equation (\ref{i1}) has the following form \cite{2}-\cite{4}
\begin{equation}\label{i2}
\frac{d^2\eta^j}{{dx^i}^2}+R^j_{ili}\eta^l=0.
\end{equation}

In the paper \cite{5} it has been shown that in the case of 
three-dimensional space
with the metrics
\begin{equation}\label{i3}
ds^2=dx^2+A(x,y,z)dy^2+2B(x,y,z)dydz+C(x,y,z)dz^2
\end{equation}
the equations of geodesic deviations
\begin{eqnarray}
\frac{d^2\eta^2}{dx^2}+R^2_{121}\eta^2+R^2_{131}\eta^3&=&0,\nonumber\\
\frac{d^2\eta^3}{dx^2}+R^3_{121}\eta^2+R^3_{131}\eta^3&=&0 \label{i4}
\end{eqnarray}
may be represented in the form of the $2\times 2$ matrix Schr\"{o}dinger
equation
\begin{eqnarray}
-\frac{d^2\eta^2}{dx^2}+(-R^2_{121}+\lambda^2)\eta^2+(-R^2_{131})\eta^3
=\lambda^2\eta^2,\nonumber\\
-\frac{d^2\eta^3}{dx^2}+(-R^3_{121})\eta^2+(-R^3_{131}+\lambda^2)\eta^3
=\lambda^2\eta^3.\label{i5}
\end{eqnarray}

On the other hand, it is known that AKNS-system \cite{6}
\begin{eqnarray}
\frac{\partial\psi_1}{\partial x}+i\lambda\psi_1&=&q(x,y,z)\psi_2,\nonumber\\
\frac{\partial\psi_2}{\partial x}-i\lambda\psi_2&=&r(x,y,z)\psi_1\label{i6}
\end{eqnarray}
can be rewritten in the form of a Schr\"{o}dinger-like equation \cite{1}
\begin{equation}\label{i7}
\left[-\left(\begin{array}{cc}1 & 0 \\ 0 & 1 \end{array}\right)
\frac{\partial^2}{\partial x^2}+\left(\begin{array}{cc} rq & q_x \\
r_x & rq\end{array}\right)\right]\begin{pmatrix} \psi_1 \\ \psi_2\end{pmatrix}
=\lambda^2\begin{pmatrix}\psi_1 \\ \psi_2\end{pmatrix}.
\end{equation}
The comparison of the systems (\ref{i7}) and (\ref{i5}) gives the following
conditions on the curvature tensor
\begin{equation}\label{i8}
{\renewcommand{\arraystretch}{1.5}
\begin{array}{rl}
\lambda^2-R^2_{121}=rq,& \lambda^2-R^3_{131}=rq,\\
R^2_{131}=-q_x,& R^3_{121}=-r_x.
\end{array}}
\end{equation}

Analogously, in the case of 4-dimensional space with a geodesic coordinate
system
\begin{equation}\label{i9}
ds^2=dt^2+g_{ab}dx^adx^b
\end{equation}
the geodesic deviations equation has the form \cite{5}
\begin{eqnarray}
\frac{d^2\eta^1}{dt^2}+R^1_{010}\eta^1+R^1_{020}\eta^2+R^1_{030}\eta^3
&=&0,\nonumber\\
\frac{d^2\eta^2}{dt^2}+R^2_{010}\eta^1+R^2_{020}\eta^2+R^2_{030}\eta^3
&=&0,\label{i10}\\
\frac{d^2\eta^3}{dt^2}+R^3_{010}\eta^1+R^3_{020}\eta^2+R^3_{030}\eta^3
&=&0.\nonumber
\end{eqnarray}

In the present paper we consider solutions of the equations (\ref{i4}) and
(\ref{i10}) obtained by the inverse scattering transform. Our consideration
is realized on the basis of a Chandrasekhar metrics \cite{7,8} (the so-called
space-time of a sufficiently general structure), which includes as 
particular cases the static and spherically symmetric solutions
(Schwarzschild and Reissner-Nordstr\"{o}m metrics), and also stationary
and axially symmetric solutions (Kerr and Kerr-Newman metrics) and so on.
In section 2 we introduce a three-dimensional analog of the 
Chandrasekhar metrics,
the particular case of which is coincide with the metrics (\ref{i3}). It is
shown that in the orthonormal basis, related with this metrics, solutions of
the system (\ref{i4}) are reduced to the solutions of the Zakharov-Shabat
problem \cite{9}. Thus, a dependence of the potential $u$ on parameters
$y$ and $z$ is described by the modified Korteweg-de Vries (mKdV) equations.
Different particular cases, in which the vector of geodesic deviation
$\boldsymbol{\eta}$ is explicitly expressed via the fundamental solutions
(Jost functions) of the Zakharov-Shabat problem, are considered at the end of
section 2. In section 3 we introduce a $3\times 3$ matrix Schr\"{o}dinger
equation which then is associated with the system of type (\ref{i10}).
Further, a dependence on parameters is reduced to evolution equations of
the well-known problem of three-wave interaction, the explicit solutions
of which was obtained by Zakharov and Manakov in 1973 \cite{10,11,12}.
It is shown that in the case of decay instability and reality of potential
matrix, the system of equations of geodesic deviation (\ref{i10}) has a
wide class of particular solutions. 
\section{Three-dimensional space}
\setcounter{equation}{0}
\subsection{The three-dimensional Chandrasekhar metrics}
Let us consider in the three-dimensional space with a signature $(-,-,-)$ a
metrics of the following form
\begin{equation}\label{e1}
\der s^{2}=-\sum_{A}e^{2\mu_{A}}(\der x^{A})^{2}-e^{2\psi}(\der x^{3}-
\sum_{A}q_{A}\der x^{A})^{2},
\end{equation}
where $A=1,2$. $\psi,\,\mu_{A}$ and $q_{A}$ are the functions on variables
$x^{1},\, x^{2},\,x^{3}$.

The orthonormal basis, related with this metrics, is defined by the
following covariant and contravariant vectors
$$e_{(1)i}=(0,\,0,\,-e^{\mu_{1}}),\quad e_{(2)i}=(0,\,-e^{\mu_{2}},\,0),$$
\begin{equation}\label{e2}
e_{(3)i}=(-e^{\psi},\,q_{1}e^{\psi},\,q_{2}e^{\psi}).
\end{equation}
$$e^{i}_{(1)}=(q_{2}e^{-\mu_{1}},\,0,\,e^{-\mu_{1}}),\quad
e^{i}_{(2)}=(q_{1}e^{-\mu_{2}},\,e^{-\mu_{2}},\,0),$$
\begin{equation}\label{e3}
e^{i}_{(3)}=(e^{-\psi},\,0,\,0).
\end{equation}
From (\ref{e2}) and (\ref{e3}) it follows that
$$e^{i}_{(a)}e_{(b)i}=\eta_{(a)(b)}=\left|
\begin{array}{ccc}
-1 & 0 & 0 \\
0 & -1 & 0 \\
0 & 0 & -1
\end{array}\right|.$$

Let
\begin{equation}\label{e4}
\bom^{A}=e^{\mu_{A}}\der x^{A},\quad \bom^{3}=e^{\psi}(\der x^{3}-
\sum_{A}q_{A}\der x^{A})
\end{equation}
be the basis 1-forms.
It is easy to see that inverse relations for (\ref{e4}) have the form
\begin{equation}\label{e5}
\der x^{A}=e^{-\mu_{A}}\bom^{A},\quad \der x^{3}=e^{-\psi}\bom^{3}+
\sum_{A}e^{-\mu_{A}}q_{A}\bom^{A}.
\end{equation}
\begin{sloppypar}
Expressing the exterior derivatives of the forms $\bom^{i}$ via the basis
2-forms $\bom^{i}\mw\bom^{j}\;(i\neq j,\,i,j=1,2,3)$, we have\end{sloppypar}
\begin{multline}
\der\bom^{A}=\sum_{B}e^{\mu_{A}}\mu_{A,B}\der x^{B}\mw\der x^{A}+
e^{\mu_{A}}\mu_{A,3}\der x^{3}\mw\der x^{A}= \\
=\sum_{B}e^{-\mu_{B}}\mu_{A,B}\bom^{B}\mw\bom^{A}+\mu_{A,3}\left[e^{-\psi}
\bom^{3}+\sum_{B}e^{-\mu_{B}}q_{B}\bom^{B}\right]\mw\bom^{A}= \\
=\sum_{B}e^{-\mu_{B}}(\mu_{A,B}+q_{B}\mu_{A,3})\bom^{B}\mw\bom^{A}+
e^{-\psi}\mu_{A,3}\bom^{3}{\scriptstyle\bigwedge}\bom^{A}.\label{e6}
\end{multline}
For the brevity of exposition let us introduce a
derivative of the function $f(x^{1},x^{2},x^{3})$ on a coordinate
$x^{A}\;(A=1,2)$ which we will denote as $f_{:A}$,
\begin{equation}\label{e7}
f_{:A}=f_{,A}+q_{A}f_{,3}.
\end{equation}
This operation is the differentiation, since it satisfies to a Leibnitz rule
$$(fg)_{:A}=fg_{:A}+gf_{:A}.$$
Using (\ref{e7}), we can rewrite
the equation (\ref{e6}) in the form
\begin{equation}\label{e8}
\der\bom^{A}=-\sum_{B}e^{-\mu_{B}}\mu_{A:B}\bom^{A}\mw\bom^{B}-
e^{-\psi}\mu_{A,3}\bom^{A}\mw\bom^{3}.
\end{equation}
In like manner we have
\begin{equation}\label{e9}
\der\bom^{3}=\sum_{A}e^{-\mu_{A}}(\psi_{:A}+q_{A,3})\bom^{A}\mw\bom^{3}-
\sum_{A,B}e^{\psi-\mu_{A}-\mu_{B}}q_{A:B}\bom^{A}\mw\bom^{B}.
\end{equation}

Further, the equations
\begin{eqnarray}
&&\frac{1}{2}T^j=\der\bom^j+\bom^j_l\mw\bom^l=\Omega^j,\\
&&\frac{1}{2}R^j_{lkm}\bom^k\mw\bom^m=\Omega^j_l
\end{eqnarray}
are called respectively the first and second Cartan structure equations,
where the Cartan 2-form $\Omega^j_l$ is
$$\Omega^j_l=\der\bom^j_l+\bom^j_k\mw\bom^k_l.$$
Owing to absence of torsion ($T^j=0$) the first Cartan structure equation gives
\begin{equation}\label{e10}
\der\bom^{3}=-\sum_{A}\bom^{3}_{A}\mw\bom^{A},
\end{equation}
\begin{equation}\label{e11}
\der\bom^{A}=-\sum_{B}\bom^{A}_{B}\mw\bom^{B}-\bom^{A}_{3}\mw\bom^{3}.
\end{equation}
These equations allow us to define the connection 1-forms $\bom^{3}_{A}$ and
$\bom^{A}_{B}$ if the forms $\der\bom^{3}$ and $\der\bom^{A}$ are known.
Since the 1-forms $\bom^{3}$ and $\bom^{A}$ are the basis forms, then
\begin{equation}\label{e12}
\bom^{i}_{j}=-\bom^{j}_{i}\quad (i,j=1,2,3).
\end{equation}
Comparing the equations (\ref{e8}) and (\ref{e9}) with the equations
(\ref{e10}) and (\ref{e11}), we obtain
\begin{eqnarray}
\bom^{3}_{A}&\!\!=\!\!&-\bom^{A}_{3}=e^{-\mu_{A}}\Psi_{A}\bom^{3}-
e^{-\psi}\mu_{A,3}\bom^{A}+ 
\frac{1}{2}\sum_{B}e^{\psi-\mu_{A}-\mu_{B}}Q_{AB}\bom^{B},\label{e13}\\
\bom^{A}_{B}&\!\!=\!\!&-\bom^{B}_{A}=-\frac{1}{2}e^{\psi-\mu_{A}-\mu_{B}}
Q_{AB}\bom^{3}
+e^{-\mu_{B}}\mu_{A:B}\bom^{A}-e^{-\mu_{A}}\mu_{B:A}\bom^{B},\label{e14}
\end{eqnarray}
where
\begin{eqnarray}
Q_{AB}&=&q_{A:B}-q_{B:A}, \\
\Psi_{A}&=&\psi_{:A}+q_{A,3}.
\end{eqnarray}
From (\ref{e13}) and (\ref{e14}) for the different connection forms we have
\begin{eqnarray}
\bom^{1}_{3}&=&e^{-\psi}\mu_{1,3}\bom^{1}-\frac{1}{2}e^{\psi-\mu_{1}-\mu_{2}}
Q_{12}\bom^{2}-e^{-\mu_{1}}\Psi_{1}\bom^{3},\nonumber \\
\bom^{2}_{3}&=&-\frac{1}{2}e^{\psi-\mu_{1}-\mu_{2}}Q_{21}\bom^{1}+
e^{-\psi}\mu_{2,3}\bom^{2}-e^{-\mu_{2}}\Psi_{2}\bom^{3},\label{e17} \\
\bom^{1}_{2}&=&e^{-\mu_{2}}\mu_{1:2}\bom^{1}-e^{-\mu_{1}}\mu_{2:1}\bom^{2}-
\frac{1}{2}e^{\psi-\mu_{1}-\mu_{2}}Q_{12}\bom^{3}.\nonumber
\end{eqnarray}

Further, in order to culculate the components of the Riemann tensor from
the second Cartan structure equation
\begin{equation}\label{e18}
\frac{1}{2}R^{i}_{jkl}\bom^{k}\mw\bom^{l}=\Omega^{i}_{j}=\der\bom^{i}_{j}+
\bom^{i}_{k}\mw\bom^{k}_{j},
\end{equation}
it is necessary at first to calculate the exterior derivatives of the connection
forms (\ref{e17}).
\begin{lem}[{\rm Chandrasekhar \cite{8}}] If $F$ is an arbitrary functions of 
the arguments
$x^{1},x^{2}$ and $x^{3}$, then
\begin{equation}\label{e19}
\der(F\bom^{3})=\sum_{A}e^{-\psi-\mu_{A}}\mathcal{D}_{A}(Fe^{\psi})\bom^{A}
\mw\bom^{3}
+\frac{1}{2}\sum_{A,B}Fe^{\psi-\mu_{A}-\mu_{B}}Q_{AB}\bom^{A}\mw\bom^{B},
\end{equation}
\begin{equation}\label{e20}
\der(F\bom^{A})=\sum_{B}e^{-\mu_{A}-\mu_{B}}(e^{\mu_{A}}F)_{:B}\bom^{B}\mw
\bom^{A}+e^{-\psi-\mu_{A}}(e^{\mu_{A}}F)_{,3}\bom^{3}\mw\bom^{A},
\end{equation}
where $\mathcal{D}_{A}$ is an operator, the action of which on an arbitrary
function $f(x^1,x^2,x^3)$ is defined by the following expression
\begin{equation}\label{e21}
\mathcal{D}_{A}f=f_{:A}+q_{A,3}f=f_{,A}+(q_{A}f)_{,3}.
\end{equation}
\end{lem}
Using this lemma, we obtain
\begin{multline}\label{e22}
\der\bom^{1}_{2}=-\sum_{A}e^{-\psi-\mu_{A}}\mathcal{D}_{A}(\frac{1}{2}
e^{2\psi-\mu_{1}-\mu_{2}}Q_{12})\bom^{A}\mw\bom^{3}-\\
-e^{-\psi-\mu_{1}}(e^{\mu_{1}-\mu_{2}}\mu_{1:2})_{,3}\bom^{1}\mw\bom^{3}+
e^{-\psi-\mu_{2}}(e^{\mu_{2}-\mu_{1}}\mu_{2:1})_{,3}\bom^{2}\mw\bom^{3}+\\
+\bom^{1}\mw\bom^{2}\left\{-\frac{1}{4}e^{2\psi-2\mu_{1}-2\mu_{2}}Q^{2}_{12}
\right.-\\
-\biggl.e^{-\mu_{1}-\mu_{2}}\left[(e^{\mu_{1}-\mu_{2}}\mu_{1:2})_{:2}+
(e^{\mu_{2}-\mu_{1}}\mu_{2:1})_{:1}\right]\biggr\},
\end{multline}
\begin{multline}\label{e23}
\bom^{1}_{3}\mw\bom^{2}_{3}=\left[e^{-2\psi}\mu_{1,3}\mu_{2,3}-\frac{1}{4}
e^{2\psi-2\mu_{1}-2\mu_{2}}Q_{12}Q_{21}\right]\bom^{1}\mw\bom^{2}-\\
-\left[e^{-\psi-\mu_{2}}\Psi_{2}\mu_{1,3}+\frac{1}{2}e^{\psi-2\mu_{1}-\mu_{2}}
\Psi_{1}Q_{21}\right]\bom^{1}\mw\bom^{3}+\\
+\left[\frac{1}{2}e^{\psi-\mu_{1}-2\mu_{2}}Q_{12}\Psi_{2}+e^{-\psi-\mu_{1}}
\Psi_{1}\mu_{2,3}\right]\bom^{2}\mw\bom^{3},
\end{multline}
\begin{multline}\label{e24}
\der\bom^{2}_{3}=-\sum_{A}e^{-\psi-\mu_{A}}\mathcal{D}_{A}(e^{\psi-\mu_{2}}
\Psi_{2})\bom^{A}\mw\bom^{3}+\\
+e^{-\psi-\mu_{1}}(\frac{1}{2}e^{\psi-\mu_{2}}Q_{21})_{,3}\bom^{1}\mw\bom^{3}
-e^{-\psi-\mu_{2}}(e^{\mu_{2}-\psi}\mu_{2,3})_{,3}\bom^{2}\mw\bom^{3}+\\
+\bom^{1}\mw\bom^{2}\left\{-\frac{1}{2}e^{\psi-\mu_{1}-2\mu_{2}}\Psi_{2}
Q_{12}+\right.\\
+\biggl.e^{-\mu_{1}-\mu_{2}}\left[(e^{\mu_{2}-\psi}\mu_{2,3})_{:1}+(\frac{1}{2}
e^{\psi-\mu_{2}}Q_{21})_{:2}\right]\biggr\},
\end{multline}
\begin{multline}\label{e25}
\bom^{1}_{2}\mw\bom^{1}_{3}=\left[e^{-\psi-\mu_{1}}\mu_{2:1}\mu_{1,3}-
\frac{1}{2}e^{\psi-\mu_{1}-2\mu_{2}}Q_{12}\mu_{1:2}\right]\bom^{1}\mw\bom^{2}+\\
+\left[\frac{1}{2}e^{-\mu_{1}-\mu_{2}}Q_{12}\mu_{1,3}-e^{-\mu_{1}-\mu_{2}}
\Psi_{1}\mu_{1:2}\right]\bom^{1}\mw\bom^{3}+\\
+\left[e^{-2\mu_{1}}\Psi_{1}\mu_{2:1}-\frac{1}{4}e^{-2\psi-2\mu_{1}-2\mu_{2}}
Q^{2}_{12}\right]\bom^{2}\mw\bom^{3},
\end{multline}
\begin{multline}\label{e26}
\der\bom^{1}_{3}=-\sum_{A}e^{-\psi-\mu_{A}}\mathcal{D}_{A}
(e^{\psi-\mu_{1}}\Psi_{1})\bom^{A}\mw\bom^{3}-\\
-e^{-\psi-\mu_{1}}(e^{\mu_{1}-\psi}\mu_{1,3})_{,3}\bom^{1}\mw\bom^{3}+
e^{-\psi-\mu_{2}}(\frac{1}{2}e^{\psi-\mu_{1}}Q_{12})_{,3}\bom^{2}\mw\bom^{3}+\\
+\bom^{1}\mw\bom^{2}\left\{-\frac{1}{2}e^{\psi-2\mu_{1}-\mu_{2}}\Psi_{1}
Q_{12}-\right.\\
-\biggl.e^{-\mu_{1}-\mu_{2}}\left[(e^{\mu_{1}-\psi}\mu_{1,3})_{:2}+
(\frac{1}{2}e^{\psi-\mu_{1}}Q_{12})_{:1}\right]\biggr\},
\end{multline}
\begin{multline}\label{e27}
\bom^{1}_{2}\mw\bom^{2}_{3}=\left[e^{-\psi-\mu_{2}}\mu_{1:2}\mu_{2,3}-
\frac{1}{2}e^{\psi-2\mu_{1}-\mu_{2}}\mu_{2:1}Q_{21}\right]\bom^{1}\mw\bom^2- \\
-\left[e^{-2\mu_{2}}\mu_{1:2}\Psi_{2}-\frac{1}{4}e^{2\psi-2\mu_{1}-2\mu_{2}}
Q_{12}Q_{21}\right]\bom^{1}\mw\bom^{3}+\\
+\left[e^{-\mu_{1}-\mu_{2}}\mu_{2:1}\Psi_{2}+\frac{1}{2}e^{-\mu_{1}-\mu_{2}}
\mu_{2,3}Q_{12}\right]\bom^{2}\mw\bom^{3}.
\end{multline}

Further, from the equations (\ref{e12}) and (\ref{e18}) we obtain
\begin{equation}\label{e28}
\frac{1}{2}R^{1}_{2kl}\bom^{k}\mw\bom^{l}=\Omega^{1}_{2}=\der\bom^{1}_{2}-
\bom^{1}_{3}\mw\bom^{2}_{3}.
\end{equation}
Substituting (\ref{e22})-(\ref{e23}) into this equation and collecting
the coefficients at $\bom^k\mw\bom^l$, we
obtain the components $R^{1}_{2kl}$ of the curvature tensor.
For example, with the object to calculate the component
$R^{1}_{212}$ we must collect the coefficients at $\bom^{1}\mw\bom^{2}$
in the expression for $\Omega^{1}_{2}$. Analogously, the components
$R^{1}_{213}$ and $R^{1}_{223}$ are obtained from $\Omega^{1}_{2}$ via the
comparison of the coefficients at $\bom^{1}\mw\bom^{3}$ and 
$\bom^{2}\mw\bom^{3}$. In like manner from the equation
\begin{equation}\label{e29}
\frac{1}{2}R^{2}_{3kl}\bom^{k}\mw\bom^{l}=\Omega^{2}_{3}=\der\bom^{2}_{3}-
\bom^{1}_{2}\mw\bom^{1}_{3}
\end{equation}
and equations (\ref{e24})-(\ref{e25}) we obtain the components
$R^{2}_{323}$ and $R^{2}_{313}$. Analogously, from equation
\begin{equation}\label{e30}
\frac{1}{2}R^{1}_{3kl}\bom^{k}\mw\bom^{l}=\Omega^{1}_{3}=\der\bom^{1}_{3}+
\bom^{1}_{2}\mw\bom^{2}_{3}
\end{equation}
and equations (\ref{e26})-(\ref{e27}) we have the component $R^{1}_{313}$.
Finally, we have the following six essential components of the curvature tensor:
\begin{multline}\label{e31}
R^{1}_{212}=-\frac{1}{4}e^{2\psi-1\mu_{1}-2\mu_{2}}Q^2_{12}-
e^{-\mu_{1}-\mu_{2}}\left[(e^{\mu_{1}-\mu_{2}}\mu_{1:2})_{:2}+
(e^{\mu_{2}-\mu_{1}}\mu_{2:1})_{:1}\right]-\\
-e^{-2\psi}\mu_{1,3}\mu_{2,3}+\frac{1}{4}e^{2\psi-2\mu_{1}-2\mu_{2}}Q_{12}
Q_{21},
\end{multline}
\begin{multline}\label{e32}
R^{1}_{213}=-e^{-\psi-\mu_{1}}\mathcal{D}_{1}(1/2e^{2\psi-\mu_{1}-
\mu_{2}}Q_{12})-e^{-\psi-\mu_{1}}(e^{\mu_{1}-\mu_{2}}\mu_{1:2})_{,3}+\\
+e^{-\psi-\mu_{2}}\Psi_{2}\mu_{1,3}+\frac{1}{2}e^{\psi-2\mu_{1}-\mu_{2}}
\Psi_{1}Q_{21},
\end{multline}
\begin{multline}\label{e33}
R^{1}_{223}=-e^{-\psi-\mu_2}\mathcal{D}_2(1/2e^{2\psi-\mu_1-\mu_2}Q_{12})
-e^{-\psi-\mu_2}(e^{\mu_2-\mu_1}\mu_{2:1})_{,3}-\\
-\frac{1}{2}e^{\psi-\mu_1-2\mu_2}Q_{12}\Psi_2-e^{-\psi-\mu_1}\Psi_1\mu_{2,3},
\end{multline}
\begin{multline}\label{e34}
R^{2}_{323}=-e^{-\psi-\mu_2}\mathcal{D}_2(e^{\psi-\mu_2}\Psi_2)-
e^{-\psi-\mu_2}(e^{\mu_2-\psi}\mu_{2,3})_{,3}-\\
-e^{-2\mu_1}\Psi_1\mu_{2:1}+\frac{1}{4}e^{2\psi-2\mu_1-2\mu_2}Q^2_{12},
\end{multline}
\begin{multline}\label{e35}
R^2_{313}=-e^{-\psi-\mu_1}\mathcal{D}_1(e^{\psi-\mu_2}\Psi_2)+
e^{-\psi-\mu_1}(1/2e^{\psi-\mu_2}Q_{21})_{,3}-\\
-e^{-\mu_1-\mu_2}\left[1/2Q_{12}\mu_{1,3}-\Psi_1\mu_{1:2}\right],
\end{multline}
\begin{multline}\label{e36}
R^1_{313}=-e^{-\psi-\mu_1}\mathcal{D}_1(e^{\psi-\mu_1}\Psi_1)-
e^{-\psi-\mu_1}(e^{\mu_1-\psi}\mu_{1,3})_{,3}-\\
-e^{-2\mu_2}\mu_{1:2}\Psi_2-\frac{1}{4}e^{2\psi-2\mu_1-2\mu_2}Q_{12}Q_{21}.
\end{multline}
\subsection{Solutions of equations of geodesic deviation
in\protect\newline the three-dimensional space}
Let us consider a particular case $(\mu_1=q_1=0)$ of the metrics (\ref{e1}).
In this case the metrics (\ref{e1}) is coincide with the 
three-dimensional metrics
considered in \cite{5} if suppose
$$A(x,y,z)=-\left(e^{2\mu_2}+q^2_2e^{2\psi}\right),
\quad B(x,y,z)=q_2e^{2\psi},$$
\begin{equation}\label{e37}
C(x,y,z)=-e^{2\psi}.
\end{equation}
At the condition $\mu_1=q_1=0$ the covariant and contravariant vectors
(\ref{e2})-(\ref{e3}) take the form
$$e_{(1)i}=(0,\,0,\,-1),\quad e_{(2)i}=(0,\,-e^{\mu_2},\,0),$$
\begin{equation}\label{e38}
e_{(3)i}=(-e^{\psi},\,0,\,q_2e^{\psi});
\end{equation}
$$e^i_{(1)}=(q_2,\,0,\,1),\quad e^i_{(2)}=(0,\,e^{-\mu_2},\,0),$$
\begin{equation}\label{e39}
e^i_{(3)}=(e^{-\psi},\,0,\,0).
\end{equation}
It is easy to see that in this orthonormal basis for the components of
the curvature tensor we have
\begin{equation}\label{e40}
R^n_{jkl}=-R_{ijkl}.
\end{equation}

It is well-known that the Riemann tensor $R_{ijkl}$ has the following
symmetry properties:
\begin{eqnarray}
R_{ijkl}&=&R_{klij},\nonumber\\
R_{ijkl}&=&-R_{jikl},\label{e41}\\
R_{ijkl}&=&-R_{ijlk}.\nonumber
\end{eqnarray}
\begin{sloppypar}
It is easy to show that the symmetry properties (\ref{e41}) decrease the
number of independent (essential) components of the Riemann tensor from $n^4$
to $n^2(n^2-1)/12$, where $n$ is a dimensionality of the space. In the case of
three-dimensional space we have six independent components of 
the curvature tensor:
$R_{1212},\,R_{1213},\,R_{1223},\,R_{1313},\,R_{2313},\,R_{2323}$.
Further, using (\ref{e40})-(\ref{e41}), we see that in the
system (\ref{i4}) among the four components of the curvature tensor only
three are independent, namely, $R^2_{121},\,R^3_{131}$ and $R^2_{131}$ 
(or $R^3_{121}$). The latter two components are coincide with each other
in virtue of (\ref{e40})-(\ref{e41}).
Therefore,\end{sloppypar}
$$R^3_{121}=-R^2_{113}.$$
Hence it immediately follows that the conditions (\ref{i8}) and the system
(\ref{i7}) are reduced to the form
\begin{equation}\label{e42}
{\renewcommand{\arraystretch}{1.5}
\begin{array}{rl}
\lambda^2-R^2_{121}=-u^2,& \lambda^2-R^3_{131}=-u^2,\\
R^3_{121}=&-R^2_{113}=u_x;
\end{array}}
\end{equation}
\begin{equation}\label{e43}
\left[-\begin{pmatrix} 1 & 0 \\ 0 & 1 \end{pmatrix}\frac{\partial^2}
{\partial x^2}+\begin{pmatrix} -u^2 & u_x \\ -u_x & -u^2 \end{pmatrix}
\right]\begin{pmatrix} \psi_1 \\ \psi_2 \end{pmatrix}=\lambda^2
\begin{pmatrix} \psi_1 \\ \psi_2 \end{pmatrix}.
\end{equation}
It is easy to see that the matrix equation (\ref{e43}) corresponds to the
Zakharov-Shabat system \cite{9}
\begin{eqnarray}
\frac{\partial\psi_1}{\partial x}+i\lambda\psi_1&=&u\psi_2,\nonumber\\
\frac{\partial\psi_2}{\partial x}-i\lambda\psi_2&=&-u\psi_1.\label{e44}
\end{eqnarray}
Thus, {\it in the orthonormal basis (\ref{e38})-(\ref{e39}), related with
the metrics (\ref{e1}), at the condition $\mu_1=q_1=0$ the AKNS-system
for the equations of geodesic deviation is 
reduced to the Zakharov-Shabat system.}
Moreover, instead the two potentials in AKNS-system we have 
now only one potential
in ZS-system.

Let us calculate the independent components of the 
curvature tensor in the system
(\ref{i4}) for the metrics (\ref{e1}) at the condition $\mu_1=q_1=0$.
From (\ref{e31}), (\ref{e32}) and (\ref{e36}) we have
\begin{eqnarray}
R^1_{212}&=&-\mu_{2,11}-\mu^2_{2,1}-\frac{1}{4}e^{-2\mu_2}q^2_{2,1}
(e^{2\psi}+1),\label{e45}\\
R^1_{213}&=&\frac{1}{2}e^{\psi-\mu_2}q_{2,11}+\frac{3}{2}e^{\psi-\mu_2}
\psi_{,1}q_{2,1}-\frac{1}{2}e^{\psi-\mu_2}\mu_{2,1}q_{2,1},\label{e46}\\
R^1_{313}&=&-\psi_{,11}-\psi^2_{,1}+\frac{1}{4}e^{2\psi-2\mu_2}q^2_{2,1}.
\label{e47}
\end{eqnarray} 

So, in the case of the metrics (\ref{e1}) our problem
of solving of the equations of geodesic deviation 
is reduced to the Zakharov-Shabat problem 
(\ref{e44}). It is known that fundamental solutions (Jost functions)
of ZS-problem are defined by the following expressions \cite{6,13,14}
\begin{equation}\label{e48}
{\renewcommand{\arraystretch}{2.0}
\begin{array}{ccl}
\varphi^-_1(x,\lambda)&=&e^{-i\lambda x}+{\displaystyle\int\limits^x_{-\infty}}
dx^\prime A_1(x,x^\prime)e^{-i\lambda x^\prime},\\
\varphi^-_2(x,\lambda)&=&{\displaystyle\int\limits^x_{-\infty}}
dx^\prime A_2(x,x^\prime)e^{-i\lambda x^\prime};
\end{array}}
\end{equation}
\begin{equation}\label{e49}
{\renewcommand{\arraystretch}{2.0}
\begin{array}{ccl}
\varphi^+_1(x,\lambda)&=&{\displaystyle\int\limits^\infty_x}
dx^\prime B_1(x,x^\prime)e^{i\lambda x^\prime},\\
\varphi^+_2(x,\lambda)&=&e^{i\lambda x}+{\displaystyle\int\limits^\infty_x}
dx^\prime
B_2(x,x^\prime)e^{i\lambda x^\prime}.
\end{array}}
\end{equation}
These solutions are linearly dependent:
\begin{eqnarray}
\varphi^-(x,\lambda)&=&c_{11}(\lambda)\varphi^+(x,\lambda)+
c_{12}(\lambda)\bar{\varphi}^+(x,\lambda),\label{e50}\\
\varphi^+(x,\lambda)&=&c_{21}(\lambda)\bar{\varphi}^-(x,\lambda)+c_{22}(\lambda)
\varphi^-(x,\lambda),\label{e51}
\end{eqnarray}
where
\begin{equation}\label{e52}
\varphi^\mp(x,\lambda)=\begin{pmatrix} \varphi^\mp_1(x,\lambda)\\ 
\varphi^\mp_2(x,\lambda)
\end{pmatrix},\quad \bar{\varphi}^\mp(x,\lambda)=\begin{pmatrix}
\varphi^\mp_2(x,-\lambda)\\
-\varphi^\mp_1(x,-\lambda)\end{pmatrix}.
\end{equation}
Further, the pair of Gel'fand-Levitan-Marchenko integral equations
can be derived
from (\ref{e50}) by means of the Fourier transform:
$$-A_2(x,y)+\Omega_L(x+y)+\int\limits^x_{-\infty}dx^\prime A_1(x,x^\prime)
\Omega_L(x^\prime+y)=0,$$
\begin{equation}\label{e54}
A_1(x,y)+\int\limits^x_{-\infty}dx^\prime A_2(x,x^\prime)\Omega_L(x^\prime
+y)=0,
\end{equation}
$$x>y.$$
Analogously, from (\ref{e51}) we have the following pair
$$B_2(x,y)+\int\limits^\infty_xdx^\prime B_1(x,x^\prime)\Omega_R(x^\prime
+y)=0,$$
\begin{equation}\label{e55}
-B_1(x,y)+\Omega_R(x+y)+\int\limits^\infty_xdx^\prime B_2(x,x^\prime)
\Omega_R(x^\prime+y)=0,
\end{equation}
$$x<y,$$
where
\begin{eqnarray}
\Omega_L&=&r_L(z)-i\sum^N_{l=1}\frac{c_{22}(\lambda_l)}{\dot{c}_{12}(\lambda_l)}
e^{-i\lambda_l z},\label{e56}\\
\Omega_R&=&r_R(z)+i\sum^N_{l=1}\frac{c_{11}(\lambda_l)}{\dot{c}_{21}(\lambda_l)}
e^{i\lambda_l z}.\label{e57}
\end{eqnarray} 
Thus, the potential $u(x)$ is expressed via the kernels $A_1,\,A_2$ and
$B_1,\,B_2$ as follows
\begin{equation}\label{e58}
{\renewcommand{\arraystretch}{2.0}
\begin{array}{lcl}
u=-2A_2(x,x),&&u=-2B_1(x,x),\\
u^2=2\frac{\displaystyle dA_1(x,x)}{\displaystyle dx},&&
u^2=-2\frac{\displaystyle dB_2(x,x)}{\displaystyle dx}.
\end{array}}
\end{equation}
In the case of a reflectionless potential $(r_L(z)=0)$ the system of
Gel'fand-Levitan-Marchenko integral equations may be solved explicitly.
In this case we obtain
\begin{equation}\label{e59}
\Omega_L=-i\sum^N_{l=1}\frac{c_{22}(\lambda_l)}{\dot{c}_{12}(\lambda_l)}
e^{-i\lambda_lz}.
\end{equation}
The system (\ref{e54}) is reduced to algebraic equations and the potential
$u(x)$ is defined by a following expression
\begin{equation}\label{e60}
u=2\frac{d}{dx}\arctan\left[\frac{\mbox{Im}\det(I-iM)}
{\mbox{Re}\det(I-iM)}\right],
\end{equation}
where
\begin{eqnarray}
M_{ij}&=&\frac{im_{Lj}}{\kappa_i+\kappa_j}e^{-i(\kappa_i+\kappa_j)x},
\label{e61}\\
m_{Lj}&=&-i\frac{c_{22}(\lambda_j)}{\dot{c}_{12}(\lambda_j)}.\nonumber
\end{eqnarray}
Here $\kappa_i$ are the poles of the transmission coefficient $T_L(\lambda)=
\frac{\displaystyle 1}{\displaystyle c_{12}(\lambda)}$. 
Analogous relations take place in the case
of system (\ref{e55}).

Let us return to the equations of geodesic deviation. From the conditions on
the curvature tensor (\ref{e42}) it follows that
\begin{eqnarray}
R^2_{121}&=&R^3_{131},\nonumber\\
R^3_{121}&=&u_x.\nonumber
\end{eqnarray}
Substituting the expressions (\ref{e45})-(\ref{e47}) into the latter
equations, we obtain
\begin{eqnarray}
\psi_{,11}+\psi^2_{,1}-\mu_{2,11}-\mu^2_{2,1}-\frac{1}{2}e^{-2\mu_2}
q^2_{2,1}(e^{2\psi}+\frac{1}{2})&=&0,\label{e62}\\
\frac{1}{2}e^{\psi-\mu_2}q_{2,11}+\frac{3}{2}e^{\psi-\mu_2}\psi_{,1}
q_{2,1}-\frac{1}{2}e^{\psi-\mu_2}\mu_{2,1}q_{2,1}&=&u_{,1}.\label{e63}
\end{eqnarray}
Thus, we have the system of differential equations (\ref{e62})-(\ref{e63})
as the conditions on the potential $u(x)$. The explicit form of $u(x)$
we will find by means of the inverse scattering problem. 
Moreover, the potential
$u(x)$ depends parametrically
on variables $y$ and $z$. According to
widely accepted methods \cite{6,13,14}, the dependence on variables $y$ and
$z$ may be represented by a nonlinear integrable equation. Indeed,
the dependence on $y$ for $\psi_1$ and $\psi_2$ from (\ref{e44}) may be
expressed in general form
\begin{eqnarray}
\psi_{1y}&=&A\psi_1+B\psi_2,\nonumber\\
\psi_{2y}&=&C\psi_1+D\psi_2.\label{e64}
\end{eqnarray}
The compatibility conditions of (\ref{e64}) with (\ref{e44}) give
(at this point, $\lambda_y=0$):
\begin{eqnarray}
A_x&=&u(B+C),\\
B_x+2i\lambda B&=&-2uA+u_y,\label{e66}\\
C_x-2i\lambda C&=&-2uA-u_y,\label{e67}
\end{eqnarray}
here $D_x=-A_x$. Further, let us suppose that
$A=\sum^3_0a_n\lambda^n,\;B=\sum^3_0b_n
\lambda^n$ and $C=\sum^3_0c_n\lambda^n$. Substituting these series into
(\ref{e66})-(\ref{e67}), we obtain for the coefficients $a_n,\,b_n$ and
$c_n$ the following expressions
\begin{eqnarray}
a_3=a_3(y),&&b_3=c_3=0,\nonumber\\
a_2=a_2(y),&&b_2=-c_2=ia_3u,\nonumber\\
a_1=-\frac{1}{2}a_3u^2,&&b_1=-\frac{1}{2}a_3u_x+ia_2u,\quad
c=-\frac{1}{2}a_3u_x-ia_2u,\nonumber\\
a_0=-\frac{1}{2}a_2u^2,&&b_0=c_0=\frac{i}{4}a_3(u_{xx}+2u^3)-\frac{1}{2}
a_2u_x.\label{e68}
\end{eqnarray}
In the equations (\ref{e66})-(\ref{e67}) the components independing on
$\lambda$ give the evolution equation $u_y=b_{0x}+2a_0u$. Using the
obtained above expressions (\ref{e68}), we obtain
for the coefficients $a_0$ and $b_0$:
$$u_y=-\frac{1}{4}ia_3(6u^2u_x+u_{xxx})-a_2(u^3+\frac{1}{2}u_{xx}).$$
Suppose $a_2=0$ and $a_3=4i$ we have the modified Korteweg-de Vries equation
\begin{equation}\label{e69}
u_y+6u^2u_x+u_{xxx}=0.
\end{equation}
Thus, the dependence on parameter $y$ for the potential $u$ is defined by the
mKdV equation. Thus, the system (\ref{e64}) has a form
\begin{eqnarray}
\psi_{1y}&=&2i\lambda(u^2-2\lambda^2)\psi_1+(4\lambda^2u+2i\lambda u_x
-2u^3-u_{xx})\psi_2,\nonumber\\
\psi_{2y}&=&(-4\lambda^2u+2i\lambda u_x+2u^3+u_{xx})\psi_1-2i\lambda
(u^2-2\lambda^2)\psi_2.\label{e70}
\end{eqnarray}
Solutions of the modified Korteweg-de Vries equation can be found
by the standard
procedure \cite{6,13,14}. When $u\rightarrow 0$ we see that the
dependence on $y$ is described
by a limiting form of the equations (\ref{e70})
\begin{eqnarray}
\psi_{1y}&=&-4i\lambda^3\psi_1,\nonumber\\
\psi_{2y}&=&4i\lambda^3\psi_2.\label{e71}
\end{eqnarray}
Let us assume that $\psi=\begin{pmatrix}\psi_1\\ \psi_2\end{pmatrix}$ is
proportional to the fundamental solution $\varphi^-$ at $x\rightarrow -\infty$.
Then, at $x\rightarrow-\infty$ we have $\psi(x,y)=f(y)\varphi^-\rightarrow
f(y)e^{-i\lambda x}\begin{pmatrix}0\\ 1\end{pmatrix}$. Substituting
$\psi_1=f(y)e^{-i\lambda x}$ into the first equation from (\ref{e71}), we
obtain $f(y)=f(0)\exp(-4i\lambda^3y)$. From (\ref{e50}) at $x\rightarrow
+\infty$ it follows that
\begin{equation}\label{e72}
\psi=f(y)\varphi^-\longrightarrow f(0)e^{-4i\lambda^3y}\left[c_{11}(\lambda,
y)e^{i\lambda x}\begin{pmatrix} 0 \\ 1\end{pmatrix}+c_{12}(\lambda,y)
e^{-i\lambda x}\begin{pmatrix} 0 \\ 1\end{pmatrix}\right].
\end{equation}
Substituting it again into (\ref{e71}), 
we obtain that $\dot{c}_{12}=0,\;\dot{c}_{11}
=8i\lambda^3$, whence
\begin{eqnarray}
c_{12}(\lambda,y)&=&c_{12}(\lambda,0),\nonumber\\
c_{11}(\lambda,y)&=&c_{11}(\lambda,0)e^{8i\lambda^3y}.\label{e73}
\end{eqnarray}
The analogous calculations for $\psi\sim\varphi^+$ give
\begin{equation}\label{e73'}
c_{22}(\lambda,y)=c_{22}(\lambda,0)e^{-8i\lambda^3y}.
\end{equation}
Using the dependence on parameter $y$ given by the relations (\ref{e73})-
(\ref{e73'}), we have for (\ref{e61}) and (\ref{e56}) the following
expressions
\begin{equation}\label{e74}
m_{Ll}(\kappa_l,y)=-i\frac{c_{22}(\kappa_l,y)}{\dot{c}_{12}(\kappa_l,0)}=
m_{Ll}(\kappa_l,0)e^{-8i\lambda^3_ly},
\end{equation}
\begin{equation}\label{e75}
\Omega_L(z,y)=\int\limits^\infty_{-\infty}\frac{d\lambda}{2\pi}R_L(\lambda,0)
e^{-8i\lambda^3y-i\lambda z}+\sum^N_{l=1}m_{Ll}(\kappa_l,0)
e^{-8i\kappa_l^3y-i\kappa_lz},
\end{equation} 
where $R_L(\lambda,0)=-c_{22}(\lambda)/c_{21}(\lambda)$. Further, from
(\ref{e58}) it follows that the potential $u(x,y)$ is expressed by the kernel
of Gel'fand-Levitan-Marchenko equations (\ref{e54}) as
$$u(x,y)=-2A_2(x,x,y).$$
In case of the reflectionless potential $(R_L(\lambda,0)=0)$ the integral
equations (\ref{e54}) are solved explicitly. In this case,
the potential $u(x,y)$ is
defined by the formula (\ref{e60}) with the matrix $M$ of the following
form
\begin{equation}\label{e76}
M_{ij}=i\frac{m_{Lj}(\kappa_j,y)}{\kappa_i+\kappa_j)}e^{-i(\kappa_i+\kappa_j)x}.
\end{equation}
In the simplest case of the one-soliton solution $(N=1)$, the matrix $M$ 
is reduced
to a scalar $M=i(m_1/2\kappa_1)\exp(-2i\kappa_1x)$. Taking into account
the relation (\ref{e74}), we see that the matrix $M$ at
$\kappa=i\lambda$ can be written as
$$M=\frac{m_1(0)}{2\lambda}e^{2\lambda x-8\lambda^3y}.$$
Thus, in case of the one-soliton solution, the potential, defined by the
expression (\ref{e60}) with the matrix (\ref{e76}), is reduced to the form
\begin{equation}\label{e77}
u(x,y)=-2\frac{\partial}{\partial x}\arctan\left[\frac{m_1(0)}{2\lambda}
e^{2\lambda x -8\lambda^3y}\right]
\end{equation}
or
\begin{equation}\label{e78}
u(x,y)=\pm 2\lambda\,\mbox{sech}\,(2\lambda x -8\lambda^3+\delta),
\end{equation}
where $\delta=\ln\left[m_1(0)/2\lambda\right]$. For $m_1(0)<0$ we take the
upper sign, for $m_1(0)>0$ the lower sign.

Supposing the analogous dependence on the parameter $z$, 
that is, defining it by the
modified Korteweg-de Vries equation of the form
\begin{equation}\label{e79}
u_z+6u^2u_x+u_{xxx}=0,
\end{equation}
we came in the case of the one-soliton solution to the following dependence
\begin{equation}\label{e80}
u(x,y,z)=\pm 2\lambda\mbox{sech}(2\lambda x -8\lambda^3y-8\lambda^3z+\delta).
\end{equation}
Thus, we see that dependence of the potential $u$ on
the parameters $y$ and $z$
 is given by the mKdV equations (\ref{e69}) and (\ref{e79}). 
Let us consider now how
the vector of geodesic deviation $\boldsymbol{\eta}$ may be expressed via
the fundamental solutions $\varphi^\mp$ of
the Zakharov-Shabat problem (\ref{e44}).
We will consider here two particular cases of the system (\ref{e62})-
(\ref{e63}).
\subsubsection{$\psi=0$}
In this case, the coefficients (\ref{e37}) of the metrics (\ref{i3}) are
\begin{equation}\label{e81}
A=-\left(e^{2\mu_2}+q^2_2\right),\quad B=q_2,\quad C=-1.
\end{equation}
And the system (\ref{e62})-(\ref{e63}) is reduced to the form
\begin{eqnarray}
\mu_{2,11}+\mu^2_{2,1}+\frac{3}{4}e^{-2\mu_2}q^2_{2,1}&=&0,\label{e82}\\
\frac{1}{2}e^{-\mu_2}q_{2,11}-\frac{1}{2}e^{-\mu_2}\mu_{2,1}q_{2,1}&=&u_{,1}.
\label{e83}
\end{eqnarray}
The latter equation, obviously, may be written as
$$\frac{1}{2}\left(e^{-\mu_2}q_{2,1}\right)_{,1}=u_{,1},$$
whence
\begin{equation}\label{e84}
u=\frac{1}{2}e^{-\mu_2}q_{2,1}.
\end{equation}
On the other hand, in virtue of (\ref{e80}) we have
\begin{equation}\label{e85}
\frac{1}{2}e^{-\mu_2}q_{2,1}=\pm 2\lambda\mbox{sech}(2\lambda x-8\lambda^3y
-8\lambda^3z+\delta).
\end{equation} 
Using the well-known relation $\sinh 2A=2\cosh A\sinh A$, we can write
(choosing the upper sign) (\ref{e85}) in the form
\begin{equation}\label{e86}
e^{-\mu_2}q_{2,1}=\frac{8\lambda\sinh(2\lambda x -8\lambda^3y-8\lambda^3z
+\delta)}{\sinh(4\lambda x-16\lambda^3y-16\lambda^3z+2\delta)}.
\end{equation}
Whence, supposing
\begin{eqnarray}
e^{-\mu_2}&=&\frac{1}{\sinh(4\lambda x-16\lambda^3y-16\lambda^3z+2\delta)},
\nonumber\\
q_{2,1}&=&8\lambda\sinh(2\lambda x-8\lambda^3y-8\lambda^3z+\delta),\nonumber
\end{eqnarray}
we obtain
\begin{eqnarray}
\mu_2&=&\ln\sinh(4\lambda x-16\lambda^3y-16\lambda^3z+2\delta),\label{e87}\\
q_2&=&4\lambda\cosh(2\lambda x-8\lambda^3y-8\lambda^3z+\delta),\label{e88}
\end{eqnarray}
or
\begin{eqnarray}
\mu_2&=&\ln\sinh(4\lambda x-16\lambda^3y-16\lambda^3z+2\delta),\label{e87'}\\
q_2&=&-4\lambda\cosh(2\lambda x-8\lambda^3y-8\lambda^z+\delta)\label{e88'}
\end{eqnarray}
in the case $m_1(0)>0$.

Thus, using (\ref{e48})-(\ref{e49}) and (\ref{e58}), we obtain that solutions
of the equations of geodesic deviation (\ref{i4}) in the case of the metrics
(\ref{e81}) are expressed via the fundamental solutions of the Zakharov-Shabat
problem as follows
\begin{eqnarray}
\eta^2&\sim&\varphi^-_1(x,\lambda)=e^{-i\lambda x}+\frac{1}{2}\int
\limits^x_{-\infty}dx^\prime\int u^2dxe^{-i\lambda x^\prime},\label{e89}\\
\eta^3&\sim&\varphi^-_2(x,\lambda)=-\frac{1}{2}\int\limits^x_{-\infty}
dx^\prime ue^{-i\lambda x^\prime}\label{e90}
\end{eqnarray}
or
\begin{eqnarray}
\eta^2&\sim&\varphi^+_1(x,\lambda)=-\frac{1}{2}\int\limits^\infty_x
dx^\prime ue^{i\lambda x^\prime},\label{e91}\\
\eta^3&\sim&\varphi^+_2(x,\lambda)=e^{i\lambda x}-\frac{1}{2}\int
\limits^\infty_xdx^\prime\int u^2dxe^{i\lambda x},\label{e92}
\end{eqnarray}
where
\begin{equation}\label{e93}
u=\frac{1}{2}e^{-\mu_2}q_{2,1},
\end{equation}
and the functions $\mu_2$ and $q_2$ are related by the equation
\begin{equation}\label{e94}
\mu_{2,11}+\mu^2_{2,1}+\frac{1}{4}e^{-2\mu_2}q^2_{2,1}=0.
\end{equation}

For example, in the case of the parameter dependence on $y$ and $z$ described
by the mKdV equations (\ref{e69}) and (\ref{e79}) at $N=1$ (one-soliton
solution) we have the following integral representations
\begin{eqnarray}
\eta^2&\sim&e^{-i\lambda x}+\lambda\int\limits^x_{-\infty}dx\tanh(2\lambda
x-8\lambda^3y-8\lambda^3z+\delta)e^{-i\lambda x},\nonumber\\
\eta^3&\sim&\mp\lambda\int\limits^x_{-\infty}dx\,\mbox{sech}(2\lambda 
x-8\lambda^3y-8\lambda^3z+\delta)e^{-i\lambda x}
\end{eqnarray}
or
\begin{eqnarray}
\eta^2&\sim&\mp\lambda\int\limits^\infty_xdx\,\mbox{sech}(2\lambda x-8\lambda^3y
-8\lambda^3z+\delta)e^{i\lambda x},\\
\eta^3&\sim&e^{i\lambda x}-\lambda\int\limits^\infty_xdx\tanh(2\lambda x
-8\lambda^3y-8\lambda^3z+\delta)e^{i\lambda x},
\end{eqnarray}
where for $m_1(0)<0$ we take the upper sign and the lower sign for $m_1(0)>0$.
The constraint (\ref{e94}) gives (here the functions $\mu_2$ and $q_2$ are
defined by (\ref{e87})-(\ref{e88}) or (\ref{e87'})-(\ref{e88'})):
$$\cosh^2(4\lambda x-16\lambda^3y-16\lambda^3z+2\delta)+3\sinh^2(2\lambda x
-8\lambda^3y-8\lambda^3z+\delta)=1.$$
\subsubsection{$\mu_2=0$}
In this case, the coefficients (\ref{e37}) of the metrics (\ref{i3}) are
\begin{equation}\label{e95}
A(x,y,z)=-\left(1+q^2_2e^{2\psi}\right),\quad B(x,y,z)=q_2e^{2\psi},
\quad C(x,y,z)=-e^{2\psi}.
\end{equation}
The system (\ref{e62})-(\ref{e63}) is reduced to the form
\begin{eqnarray}
&&\psi_{,11}+\psi^2_{,1}-\frac{1}{2}q^2_{2,1}(e^{2\psi}+\frac{1}{2})=0,
\label{e96}\\
&&\frac{1}{2}e^{\psi}q_{2,11}+\frac{3}{2}e^{\psi}\psi_{,1}q_{2,1}=u_{,1}.
\label{e97}
\end{eqnarray}
Using the substitution $\theta=\frac{\displaystyle\psi}{\displaystyle 3}$,
we obtain from the latter equation
\begin{equation}\label{e98}
\frac{1}{2}\left(e^{\theta/3}q_{2,1}\right)_{,1}=u_{,1}.
\end{equation}
Therefore, in this case the vector of geodesic deviation $\boldsymbol{\eta}$
is also expressed via the fundamental solutions $\varphi^\mp(x,\lambda)$
of the form (\ref{e89})-(\ref{e90}) or (\ref{e91})-(\ref{e92}). 
At this point,
\begin{equation}\label{e99}
u=\frac{1}{2}e^{\theta/3}q_{2,1}
\end{equation}
and the functions $\theta$ and $q_2$ are related by the equation
\begin{equation}\label{e100}
\theta_{,11}+\frac{1}{3}\theta^2_{,1}-\frac{3}{2}q_{2,1}(e^{2/3\theta}+
\frac{1}{2})=0.
\end{equation}
In the case of the one-soliton solution, a potential $u$
is defined by the expression
(\ref{e80}), and the functions $\theta$ and $q_2$ are respectively equal to
\begin{eqnarray}
\theta&=&3\ln\mbox{csch}(4\lambda x-16\lambda^3y-16\lambda^3z+2\delta),\\
q_2&=&\pm\cosh(2\lambda x-8\lambda^3y-8\lambda^3z+\delta).
\end{eqnarray}

More complicated case $\mu_2\neq 0,\;\psi\neq 0$ and also multi-soliton
solutions will be considered in a separate paper.
%
\section{Four-dimensional space}
\setcounter{equation}{0}
\subsection{$3\times 3$ matrix Schr\"{o}dinger equation}
Let us consider the following linear problem
\begin{equation}\label{f1}
\psi_{,4}=i\zeta D\psi+N\psi,
\end{equation}
where $\psi_{,4}=\frac{\displaystyle\partial}{\displaystyle\partial x^4}
\psi,\;x^4=it;\;\zeta$ is a spectral parameter and $\psi$ is a
$3\times 1$ matrix (vector) of the form
$$\psi=\begin{pmatrix}
\psi_1\\
\psi_2\\
\psi_3
\end{pmatrix}.$$
The $3\times 3$ matrices $D$ and $N$ (a potential matrix) are
$$D=\begin{pmatrix}
\mp d_1 & 0 & 0\\
0 & \pm d_2 & 0\\
0 & 0 & \mp d_3
\end{pmatrix},\quad N=\begin{pmatrix}
0 & N_{12} & N_{13}\\
N_{21} & 0 & N_{23}\\
N_{31} & N_{32} & 0
\end{pmatrix}.$$
The system (\ref{f1}) may be rewritten (see Appendix) in the following form
($3\times 3$ matrix Schr\"{o}dinger equation)
\begin{equation}\label{f2}
-I\psi_{,44}+\mathfrak{N}\psi=d^2\zeta^2\psi,
\end{equation}
where $I$ is a $3\times 3$ unit matrix, $d=(d_1,\,d_2,\,d_3)$ and
$$\mathfrak{N}=
\begin{pmatrix}
 N_{12}N_{21}+N_{13}N_{31}& N_{12,4}+N_{13}N_{32}& N_{13,4}+N_{12}N_{23}\\
 N_{21,4}+N_{23}N_{31} & N_{21}N_{12}+N_{23}N_{32} & N_{23,4}+N_{21}N_{13}\\
 N_{31,4}+N_{32}N_{21} & N_{32,4}+N_{31}N_{12} & N_{31}N_{13}+N_{32}N_{23}
\end{pmatrix}.$$
     
Further, it is easy to see that for the metrics (\ref{i9}) the geodesic
deviation equation (\ref{i10}),
\begin{eqnarray}     
\eta^1_{,44}+R^1_{414}\eta^1+R^1_{424}\eta^2+R^1_{434}\eta^3&=&0,\nonumber\\
\eta^2_{,44}+R^2_{414}\eta^1+R^2_{424}\eta^2+R^2_{434}\eta^3&=&0,\label{f3}\\
\eta^3_{,44}+R^3_{414}\eta^1+R^3_{424}\eta^2+R^3_{434}\eta^3&=&0,\nonumber
\end{eqnarray}
can be rewritten in the form of the $3\times 3$ matrix Schr\"{o}dinger operator
\begin{eqnarray}
-\eta^1_{,44}+(-R^1_{414}+d^2_1\zeta^2)\eta^1+(-R^1_{424})\eta^2+(-R^1_{434})
\eta^3&=&d^2_1\zeta^2\eta^1,\nonumber\\
-\eta^2_{,44}+(-R^2_{414})\eta^1+(-R^2_{242}+d^2_2\zeta^2)\eta^2+(-R^2_{434})
\eta^3&=&d^2_2\zeta^2\eta^2,\label{f4}\\
-\eta^3_{,44}+(-R^3_{414})\eta^1+(-R^3_{424})\eta^2+(-R^3_{434}+d^2_3\zeta^2)
\eta^3&=&d^2_3\zeta^2\eta^3.\nonumber
\end{eqnarray}
Comparing these equations with equations (\ref{f2}), we obtain the following
conditions on the curvature tensor:
\begin{eqnarray}
&&\phantom{-}d^2_1\zeta^2-R^1_{414}=N_{12}N_{21}+N_{13}N_{31},\nonumber\\
&&\phantom{-}d^2_2\zeta^2-R^2_{424}=N_{21}N_{12}+N_{23}N_{32},\nonumber\\
&&\phantom{-}d^2_3\zeta^2-R^3_{434}=N_{31}N_{13}+N_{32}N_{23},\nonumber\\
&&-R^1_{424}=N_{12,4}+N_{13}N_{32},\nonumber\\
&&-R^1_{434}=N_{13,4}+N_{12}N_{23},\nonumber\\
&&-R^2_{414}=N_{21,4}+N_{23}N_{31},\nonumber\\
&&-R^2_{434}=N_{23,4}+N_{21}N_{13},\nonumber\\
&&-R^3_{414}=N_{31,4}+N_{32}N_{21},\nonumber\\
&&-R^3_{414}=N_{31,4}+N_{32}N_{21}.\label{f5}
\end{eqnarray}
\subsection{Chandrasekhar metrics}
In the 4-dimensional space with a signature $(-,-,-,-)$ the Chandrasekhar
metrics is defined by the following expression \cite{8} 
\begin{equation}\label{f7}
\der s^{2}=-\sum_{A}e^{2\mu_{A}}(\der x^{A})^{2}-e^{2\psi}(\der x^{1}-
\sum_{A}q_{A}\der x^{A})^{2},
\end{equation}
where $A=2,3,4$. $\psi,\,\mu_{A}$ and $q_{A}$ are the functions on variables
$x^{1},\, x^{2},\,x^{3},\,x^{4}$.

The orthonormal tetrad, related with the metrics
(\ref{f7}), is defined by the following covariant  
basis vectors:
\begin{eqnarray}
e_{(4)i}=(-e^{\mu_{4}},\,0,\,0,\,0), && e_{(1)i}=(q_{4}e^{\psi},\,-e^{\psi},\,
q_{2}e^{\psi},\,q_{3}e^{\psi}),\nonumber \\
e_{(2)i}=(0,\,0,\,-e^{\mu_{2}},\,0), && e_{(3)i}=(0,\,0,\,0,\,-e^{\mu_{3}}).
\label{f8}
\end{eqnarray}
And also the contravariant basis vectors are
\begin{eqnarray}
e^{i}_{(4)}=(e^{-\mu_{4}},\,q_{4}e^{-\mu_{4}},\,0,\,0), && e^{i}_{(1)}=(0,
\,e^{-\psi},\,
0,\,0),\nonumber \\
e^{i}_{(2)}=(0,\,q_{2}e^{-\mu_{2}},\,e^{-\mu_{2}},\,0), && e^{i}_{(3)}=(0,\,
q_{3}e^{-\mu_{3}},\,0,\,e^{-\mu_{3}}).\label{f9}
\end{eqnarray}
From (\ref{f8}) and (\ref{f9}) it is easy to see that
$$e^{i}_{(a)}e_{(b)i}=\eta_{(a)(b)}=\left|
\begin{array}{cccc}
-1 & 0 & 0 & 0 \\
0 & -1 & 0 & 0 \\
0 & 0 & -1 & 0 \\
0 & 0 & 0 & -1
\end{array}\right|.$$
Therefore, in this orthonormal basis for the components of the curvature tensor
we have always
\begin{equation}\label{f10}
R^{n}_{klm}=-R_{iklm}.
\end{equation}

Moreover, at $\der t=-i\der x^4,\;\nu=\mu_4$ and $\omega=iq_4$ there exists
an analytic continuation of the basis (\ref{f8})-(\ref{f9}) with the signature
$(-,\,-,\,-,\,-)$ onto a basis with a signature $(-,\,-,\,-,\,+)$, the
covariant and contravariant vectors of which have the form
\begin{equation}
{\renewcommand{\arraystretch}{1.3}
\begin{array}{lcl}
e_{(1)i}=(\omega e^\psi,\,-e^\psi,\,q_2e^\psi,\,q_3e^\psi), &&
e_{(2)i}=(0,\,0,\,-e^{\mu_2},\,0),\\
e_{(3)i}=(0,\,0,\,0,\,-e^{\mu_3}), &&
e_{(4)i}=(e^\nu,\,0,\,0,\,0).\\
e^i_{(1)}=(0,\,e^{-\psi},\,0,\,0), &&
e^i_{(2)}=(0,\,q_2e^{-\mu_2},\,e^{-\mu_2},\,0),\\
e^i_{(3)}=(0,\,q_3e^{-\mu_3},\,0,\,e^{-\mu_3}), &&
e^i_{(4)}=(e^{-\nu},\,\omega e^{-\nu},\,0,\,0).
\end{array}}
\end{equation}

It is obvious that in the orthonormal basis (\ref{f8})-(\ref{f9}) among
the nine components of the curvature tensor of the system (\ref{f3}) 
only six are independent, namely,
$$R^1_{414},\;R^2_{424},\;R^3_{434},\;R^1_{424},\;R^1_{434},\;R^2_{434}.$$
In the orthonormal basis (\ref{f8})-(\ref{f9}) for the metrics (\ref{f7})
these components have the form \cite{8}
\begin{multline}\label{f13}
-R_{1414}=-e^{-\psi-\mu_4}\mathcal{D}_4\left(e^{\psi-\mu_4}\Psi_4\right)-
e^{-2\mu_2}\Psi_2\mu_{4:2}-e^{-2\mu_3}\Psi_3\mu_{4:3}-\\
-e^{-\psi-\mu_4}\left(e^{\mu_4-\psi}\mu_{4,1}\right)_{,1}+\frac{1}{4}
e^{2\psi-2\mu_4}\left[e^{-2\mu_2}Q^2_{24}+e^{-2\mu_3}Q^2_{34}\right],
\end{multline}
\begin{multline}\label{f14}
-R_{2424}=-e^{-\mu_2-\mu_4}\left[\left(e^{\mu_2-\mu_4}\mu_{2:4}\right)_{:4}
+\left(e^{\mu_4-\mu_2}\mu_{4:2}\right)_{:2}\right]-\\
-e^{-2\mu_3}\mu_{4:3}\mu_{2:3}-\frac{3}{4}e^{2\psi-2\mu_2-2\mu_4}Q^2_{24}
-e^{-2\psi}\mu_{2,1}\mu_{4,1},
\end{multline}
\begin{multline}\label{f15}
-R_{3434}=-e^{-\mu_3-\mu_4}\left[\left(e^{\mu_3-\mu_4}\mu_{3:4}\right)_{:4}
+\left(e^{\mu_4-\mu_3}\mu_{4:3}\right)_{:3}\right]-\\
-e^{-2\mu_2}\mu_{4:2}\mu_{3:2}-\frac{3}{4}e^{2\psi-2\mu_3-2\mu_4}Q^2_{34}
-e^{-2\psi}\mu_{3,1}\mu_{4,1},
\end{multline}
\begin{multline}\label{f16}
R_{1424}=e^{\psi-2\mu_4-\mu_2}Q_{42}\left(\Psi_4-\frac{1}{2}\mu_{2:4}\right)
+\frac{1}{2}e^{-\mu_4-\mu_2}\left(e^{\psi-\mu_4}Q_{42}\right)_{:4}+\\
+\frac{1}{2}e^{\psi-2\mu_3-\mu_2}Q_{32}\mu_{4:3}+e^{-\mu_4-\mu_2}\left(
e^{-\psi+\mu_4}\mu_{4,1}\right)_{:2}-e^{-\psi-\mu_2}\mu_{2,1}\mu_{4:2},
\end{multline}
\begin{multline}\label{f17}
R_{1434}=e^{\psi-2\mu_4-\mu_3}Q_{43}\left(\Psi_4-\frac{1}{2}\mu_{3:4}\right)
+\frac{1}{2}e^{-\mu_4-\mu_3}\left(e^{\psi-\mu_4}Q_{43}\right)_{:4}+\\
+\frac{1}{2}e^{\psi-2\mu_2-\mu_3}Q_{23}\mu_{4:2}+e^{-\mu_4-\mu_3}\left(
e^{-\psi+\mu_4}\mu_{4,1}\right)_{:3}-e^{-\psi-\mu_3}\mu_{3,1}\mu_{4:3},
\end{multline}
\begin{multline}\label{f18}
R_{2434}=e^{-\mu_2-\mu_3}\left[\mu_{4:32}+\mu_{4:3}\left(\mu_4-\mu_3\right)
_{:2}-\mu_{4:2}\mu_{2:3}\right]-\\
-\frac{3}{4}e^{2\psi-\mu_3-2\mu_4-\mu_2}Q_{34}Q_{42}-\frac{1}{2}e^{-\mu_3-
\mu_2}Q_{32}\mu_{4,1},
\end{multline}
where
\begin{eqnarray}
\mathcal{D}_Af&=&f_{,A}+\left(q_Af\right)_{,1},\nonumber\\
\Psi_A&=&\psi_{:A}+q_{A,1},\nonumber\\
Q_{AB}&=&q_{A:B}-q_{B:A},\nonumber\\
f_{:A}&=&f_{,A}+q_{A}f_{,1}.\nonumber
\end{eqnarray}
\subsection{Solutions of equations of geodesic deviation
in\protect\newline the four-dimensional space}
It is easy to see that the Chandrasekhar metrics (\ref{f7}) coincides with the
metrics (\ref{i9}) at $\mu_4=q_4=0$. In this case, the orthonormal basis
is reduced to the form
$$
{\renewcommand{\arraystretch}{1.3}
\begin{array}{lcl}
e_{(1)i}=(0,\,-e^\psi,\,q_2e^\psi,\,q_3e^\psi),&&
e_{(2)i}=(0,\,0,\,-e^{\mu_2},\,0),\\
e_{(3)i}=(0,\,0,\,0,\,-e^{\mu_3}),&&
e_{(4)i}=(-1,\,0,\,0,\,0);\\
e^i_{(1)}=(0,\,e^{-\psi},\,0,\,e^{-\mu_3}),&&
e^i_{(2)}=(0,\,q_2e^{-\mu_2},\,e^{-\mu_2},\,0),\\
e^i_{(3)}=(0,\,q_3e^{-\mu_3},\,0,\,e^{-\mu_3}),&&
e^i_{(4)}=(1,\,0,\,0,\,0).
\end{array}}$$
It is obvious that in this basis we have $R^n_{jkl}=-R_{ijkl}$, and the
components of the curvature tensor (\ref{f13})-(\ref{f18}) are
\begin{eqnarray}
-R_{1414}&=&-\psi_{,44}-\psi^2_{,4}+\frac{1}{4}e^{2\psi}\left[e^{-2\mu_2}
q^2_{2,4}+e^{-2\mu_3}q^2_{3,4}\right],\label{f18'}\\
-R_{2424}&=&-\mu_{2,44}-\mu^2_{2,4}-\frac{3}{4}e^{2\psi-2\mu_2}q^2_{2,4},\\
-R_{3434}&=&-\mu_{3,44}-\mu^2_{3,4}-\frac{3}{4}e^{2\psi-2\mu_3}q^2_{3,4},\\
R_{1424}&=&-\frac{1}{2}e^{\psi-\mu_2}\left[q_{2,44}-\mu_{2,4}q_{2,4}+
3\psi_{,4}q_{2,4}\right],\\
R_{1434}&=&-\frac{1}{2}e^{\psi-\mu_3}\left[q_{3,44}-\mu_{3,4}q_{3,4}+
3\psi_{,4}q_{3,4}\right],\\
R_{2434}&=&\frac{3}{4}e^{2\psi-\mu_2-\mu_3}q_{2,4}q_{3,4}.\label{f19}
\end{eqnarray}

Further, let us define now the evolution equations related with the problem
(\ref{f1}). We consider the following system
\begin{eqnarray}
\psi_{,4}&=&i\zeta D\psi+N\psi,\nonumber\\
\psi_{,1}&=&Q\psi,\nonumber
\end{eqnarray}
where $\psi_{,1}=\frac{\displaystyle\partial}{\displaystyle\partial x^1},\;
x^1$ is a parameter of the considered problem. $D,\,N,\,Q$ be the $3\times 3$
matrices. At this point, $D$ is diagonal: $D=d_i\delta_{ij},\;d_i=const$;
$N$ is such a matrix that $N_{ii}=0$. From the compatibility condition
$\psi_{,14}=\psi_{,41}$ and the requirement $\zeta_{,1}=0$ we obtain
$$Q_{,4}=N_{,1}+i\zeta[D,Q]+[N,Q].$$
Decomposing $Q$ in the form
$$Q=Q^{(1)}\zeta+Q^{(0)},$$
we have $Q^{(0)}_{,4}=N_{,1}+[N,Q^{(0)}]$, 
whence we obtain the
system of $n(n-1)$ equations (see \cite{6}):
\begin{equation}\label{f20}
N_{lj,1}-a_{lj}N_{lj,4}=\sum_k(a_{lk}-a_{kj})N_{lk}N_{kj},
\end{equation}
where
$$a_{lj}=\frac{1}{i}\frac{q_l-q_j}{d_l-d_j}=a_{jl}.$$
Equations (\ref{f20}) may be reduced to the standard system of nonlinear
equations of three-wave interaction. Namely, we obtain
\begin{eqnarray}
Q_{1,1}+C_1Q_{1,4}&=&i\gamma_1Q^\ast_2Q^\ast_3,\nonumber\\
Q_{2,1}+C_2Q_{2,4}&=&i\gamma_2Q^\ast_1Q^\ast_3,\label{f21}\\
Q_{3,1}+C_3Q_{3,4}&=&i\gamma_3Q^\ast_1Q^\ast_2,\nonumber
\end{eqnarray}
where $\gamma_1\gamma_2\gamma_3=-1$, $\gamma_i=\pm 1$ and
\begin{equation}\label{f22}
{\renewcommand{\arraystretch}{1.5}
\begin{array}{lcl}
N_{12}=-iQ_3/\sqrt{\beta_{13}\beta_{23}},&&
N_{31}=-iQ_2/\sqrt{\beta_{12}\beta_{23}},\\
N_{23}=+iQ_1/\sqrt{\beta_{12}\beta_{13}},&&
N_{13}=-\gamma_1\gamma_3N^\ast_{31},\\
N_{32}=\gamma_2\gamma_3N^\ast_{23},&&
N_{21}=\gamma_1\gamma_2N^\ast_{12},
\end{array}}
\end{equation}
here
$$q_j=-i\frac{C_1C_2C_3}{C_j},\quad\beta_{lj}=d_l-d_j=C_j-C_l,$$
$$C_3>C_2>C_1.$$
In the system (\ref{f21}) there is a decay instability (for the waves with
positively defined energy) if the sign of one $\gamma_n$ is different from
the other, 
and also there is an explosive instability when $\gamma_1=\gamma_2=\gamma_3=-1$.
Solutions of the system (\ref{f21}) was obtained by Zakharov and Manakov in
1973 \cite{10,11,12}. They have the form
\begin{multline}\label{f23}
Q_1=\sqrt{\beta_{12}\beta_{13}}\frac{2\chi_3}{\mathfrak{D}}e^{i\xi_3(x^4-C_1x^1-
\bar{\varphi}_1)}\biggl[e^{\chi_1(x^4-C_3x^1-\varphi_3}-\biggr.\\
-\biggl.\gamma_1\gamma_2\frac{\bar{\zeta}_1-\bar{\zeta}_3}{\zeta^\ast_1-
\zeta_3}e^{-\chi_1(x^4-C_3x^1-\varphi_3)}\biggr],
\end{multline}
\begin{equation}\label{f24}\!\!\!\!\!\!
Q_2=\frac{-4\chi_1\chi_3\beta_{13}\gamma_2\gamma_3}{\sqrt{\beta_{12}\beta_{23}}
(\bar{\zeta}_1-\zeta^\ast_3)\mathfrak{D}}e^{-i\xi_1(x^4-C_3x^1-\bar{\varphi}_3)}
e^{-i\xi_3(x^4-C_1x^1-\bar{\varphi}_1)},
\end{equation}
\begin{multline}\label{f25}
Q_3=\sqrt{\beta_{13}\beta_{23}}\gamma_1\gamma_2\frac{2\chi_1}{\mathfrak{D}}
e^{i\xi_1(x^4-C_3x^1-\bar{\varphi}_3)}\biggl[e^{\chi_3(x^4-C_1x^1-\varphi_1)}
-\biggr.\\
-\biggl.\gamma_2\gamma_3\frac{\bar{\zeta}^\ast_1-\zeta^\ast_3}
{\bar{\zeta}^\ast_1-\zeta_3}e^{-\chi_3(x^4-C_1x^1-\varphi_1)}\biggr],
\end{multline}
where
\begin{multline}
\mathfrak{D}=\left[e^{\chi_1(x^4-C_3x^1-\varphi_3)}-\gamma_1
\gamma_2e^{-\chi_1(x^4-
C_3x^1-\varphi_3)}\right]\times\\
\times\left[e^{\chi_3(x^4-C_1x^1-\varphi_1)}-\gamma_2\gamma_3e^{-\chi_3(x^4-
C_1x^1-\varphi_1)}\right]+\\
+\gamma_1\gamma_3\frac{\left(\bar{\zeta}_1-\bar{\zeta}^\ast_1\right)\left(
\zeta_3-\zeta^\ast_3\right)}{\left(\bar{\zeta}_1-\zeta^\ast_3\right)\left(
\bar{\zeta}^\ast_1-\zeta_3\right)}e^{-\chi_1(x^4-C_3x^1-\varphi_3)}
e^{-\chi_3(x^4-C_1x^1-\varphi_1)},
\end{multline}
$$\bar{\zeta}_1=\frac{\xi_1-i\chi_1}{\beta_{12}},\quad\zeta_3=\frac{\xi_3-
i\chi_3}{\beta_{23}}.$$

Supposing now that the matrix $N$ is real ($N^\ast=N$) and choosing $\gamma_1=
\gamma_3=1,\;\gamma_2=-1$, we obtain from (\ref{f22})
\begin{equation}\label{f26}
{\renewcommand{\arraystretch}{1.5}
\begin{array}{lcl}
N_{12}=-\mbox{Re}(iQ_3/\sqrt{\beta_{13}\beta_{23}}),&&
N_{31}=-\mbox{Re}(iQ_2/\sqrt{\beta_{12}\beta_{23}}),\\
N_{23}=+\mbox{Re}(iQ_1/\sqrt{\beta_{12}\beta_{13}}),&&
N_{13}=-N_{31},\\
N_{32}=-N_{23},&&
N_{21}=-N_{12}.
\end{array}}
\end{equation}
It is obvious that the latter three conditions in (\ref{f26}) are equivalent
to antisymmetry of the matrix $N$.

Thus, we assume that the potential matrix $N$ is real and antisymmetric.
Taking it into account and also the expressions (\ref{f18'})-(\ref{f19}),
we obtain from the conditions on the curvature tensor (\ref{f5}) the following
system of differential equations:
\begin{eqnarray}
&&-\psi_{,44}-\psi^2_{,4}+\frac{1}{4}e^{2\psi}\left[e^{-2\mu_2}q^2_{2,4}+
e^{-2\mu_3}q^2_{3,4}\right]=N^2_{12}+N^2_{13}+d^2_1\zeta^2,\label{f27}\\
&&-\mu_{2,44}-\mu^2_{2,4}-\frac{3}{4}e^{2\psi-2\mu_2}q^2_{2,4}=N^2_{12}+
N^2_{23}+d^2_2\zeta^2,\\
&&-\mu_{3,44}-\mu^2_{3,4}-\frac{3}{4}e^{2\psi-2\mu_3}q^2_{3,4}=N^2_{13}+
N^2_{23}+d^2_3\zeta^2,\\
&&-\frac{1}{2}e^{\psi-\mu_2}\left[q_{2,44}-\mu_{2,4}q_{2,4}+3\psi_{,4}q_{2,4}
\right]=N_{23}N_{31},\\
&&-\frac{1}{2}e^{\psi-\mu_3}\left[q_{3,44}-\mu_{3,4}q_{3,4}+3\psi_{,4}q_{3,4}
\right]=N_{12}N_{23},\\
&&\phantom{-}\frac{3}{4}e^{2\psi-\mu_2-\mu_3}q_{2,4}q_{3,4}=N_{31}N_{12}.
\label{f32}
\end{eqnarray}
Obviously, this system has a great number of particular cases. For example,
let us consider one simplest case.
\subsubsection{$q_3=\text{const},\;\psi=\mu_2=0$}
In this case for the metrics (\ref{i9}) we have
$$
{\renewcommand{\arraystretch}{1.5}
\begin{array}{lll}
g_{11}=-1,&\quad g_{22}=-(1+q^2_2),&\quad g_{33}=-(e^{2\mu_3}+\text{const}^2),
\\
g_{12}=2q_2,&\quad g_{13}=2\text{const},&\quad g_{23}=-2\text{const}q_2.
\end{array}}
$$
The system (\ref{f27})-(\ref{f32}) is reduced to the form
\begin{eqnarray}
&&\phantom{-}q^2_{2,4}=N^2_{12}+N^2_{13}+d^2_1\zeta^2,\label{f33}\\
&&-\frac{3}{4}q^2_{2,4}=N^2_{12}+N^2_{23}+d^1_2\zeta^2,\\
&&-\mu_{3,44}-\mu^2_{3,4}=N^2_{13}+N^2_{23}+d^2_3\zeta^2,\\
&&-\frac{1}{2}q_{2,44}=N_{23}N_{31},\\
&&\phantom{-}N_{12}N_{23}=0,\\
&&\phantom{-}N_{31}N_{12}=0.\label{f38}
\end{eqnarray}
From the latter two equations it follows that $N_{12}=0$. 
Therefore, in this case, the
potential matrix $N$ has a form
$$\begin{pmatrix}
0 & 0 & -N_{31}\\
0 & 0 & N_{23}\\
N_{31} & -N_{23} & 0
\end{pmatrix}.$$
Further, at $d^2_1=-4/3d^2_2$ from (\ref{f33})-(\ref{f38}) it
follows that
\begin{eqnarray}
&&-\frac{1}{2}q_{2,44}=N_{23}N_{31},\label{f39}\\
&&-\mu_{3,44}-\mu^2_{3,4}=-\frac{1}{3}N^2_{23}+d^2_3\zeta^2,\\
&&\phantom{-}N^2_{13}+\frac{4}{3}N^2_{23}=0.\label{f41}
\end{eqnarray}
In accordance with (\ref{f26}), the components $N_{23}$ and
$N_{31}$ are defined as
\begin{multline}
N_{23}=\frac{2\chi_3}{\mathfrak{D}}\Biggl[\frac{2(\beta_{23}\chi_1-\beta_{12}
\chi_3)}{(\beta_{23}\xi_1-\beta_{12}\xi_3)^2+(\beta_{23}\chi_1-\beta_{12}
\chi_3)^2}\Biggr.\times\\
\times\cos\xi_3\left(x^4-C_1x^1-\bar{\varphi}_1\right)e^{-\chi_1\left(x^4-
C_3x^1-\varphi_3\right)}-\\
-\Biggl.2\sin\xi_3\left(x^4-C_1x^1-\bar{\varphi}_1\right)\cosh\chi_1\left(
x^4-C_3x^1-\varphi_3\right)\Biggr].
\end{multline}
\begin{multline}
N_{31}=-\frac{4\chi_1\chi_3\beta_{13}(\beta_{23}\xi_1-\beta_{12}\xi_3)}
{(\beta_{23}\xi_1-\beta_{12}\xi_3)^2+(\beta_{23}\chi_1-\beta_{12}\chi_3)^2
\mathfrak{D}}\times\\
\times\Biggl[\sin\xi_1\left(x^4-C_3x^1-\bar{\varphi}_3\right)\cos\xi_3\left(
x^4-C_1x^1-\bar{\varphi}_1\right)+\Biggr.\\
+\Biggl.\cos\xi_1\left(x^4-C_3x^1-\bar{\varphi}_3\right)\sin\xi_3\left(
x^4-C_1x^1-\bar{\varphi}_1\right)\Biggr],
\end{multline}
where
\begin{multline}
\mathfrak{D}=4\cosh\chi_1\left(x^4-C_3x^1-\varphi_3\right)\cosh\chi_3\left(
x^4-C_1x^1-\varphi_1\right)+\\
+\frac{4\chi_1\chi_3}{\beta^2_{12}(\xi^2_3+\chi^2_3)+2\beta_{12}\beta_{23}
(\xi_1\xi_3+\chi_1\chi_3)+\beta^2_{23}(\xi^2_1+\chi^2_1)}\times\\
\times e^{-\chi_1\left(x^4-C_3x^1-\varphi_3\right)}e^{-\chi_3\left(x^4-C_1x^1
-\varphi_1\right)}.
\end{multline}
After very cumbersome but elementary calculations it is easy to verify
that solutions of the system (\ref{f39})-(\ref{f41}) exist.
\section*{Appendix}
\setcounter{equation}{0}
\renewcommand{\theequation}{A.\arabic{equation}}
Let us consider the system (\ref{f1}) with the matrix
$$D=\begin{pmatrix}
-d_1 & 0 & 0\\
0 & d_2 & 0\\
0 & 0 & -d_3\\
\end{pmatrix}.$$
Differentiating (\ref{f1}) and excluding the first derivatives $\psi_{,4}$,
we obtain the following system
\begin{multline}\label{a1}
-\psi_{1,44}+(N_{12}N_{21}+N_{13}N_{31})\psi_1+(N_{12,4}+N_{13}N_{32}
-i\zeta d_1N_{12}+i\zeta d_2N_{12})\psi_2+\\
+(N_{13,4}+N_{12}N_{23}-i\zeta d_1N_{13}-i\zeta d_3N_{13})\psi_3=
\zeta^2d^2_1\psi_1,
\end{multline}
\begin{multline}
-\psi_{2,44}+(N_{21,4}+N_{23}N_{31}+i\zeta d_2N_{21}-i\zeta d_1N_{21})\psi_1
+(N_{21}N_{12}+N_{23}N_{32})\psi_2+\\
+(N_{23,4}+N_{21}N_{13}+i\zeta d_2N_{23}-i\zeta d_3N_{23})\psi_3=
\zeta^2d^2_2\psi_2,
\end{multline}
\begin{multline}\label{a3}
-\psi_{3,44}+(N_{31,4}+N_{32}N_{21}-i\zeta d_3N_{31}-i\zeta d_1N_{31})\psi_1
+(N_{32,4}+N_{31}N_{12}-i\zeta d_3N_{32}+\\
+i\zeta d_2N_{32})\psi_2+(N_{31}N_{13}+N_{32}N_{23})\psi_3=\zeta^2d^2_3\psi_3.
\end{multline}
Analogously, for the matrix
$$D=\begin{pmatrix}
d_1 & 0 & 0\\
0 & -d_2 & 0\\
0 & 0 & d_3
\end{pmatrix}$$
we have the system
\begin{multline}\label{a4}
-\psi_{1,44}+(N_{12}N_{21}+N_{13}N_{31})\psi_1+
(N_{12,4}+N_{13}N_{32}+i\zeta d_1N_{12}-i\zeta d_2N_{12})\psi_2+\\
+(N_{13,4}+N_{12}N_{23}+i\zeta d_1N_{13}+i\zeta d_3N_{13})\psi_3=
\zeta^2d^2_1\psi_1,
\end{multline}
\begin{multline}
-\psi_{2,44}+(N_{21,4}+N_{23}N_{31}-i\zeta d_2N_{21}+i\zeta d_1N_{21})\psi_1+
(N_{21}N_{12}+N_{23}N_{32})\psi_2+\\
+(N_{23,4}+N_{21}N_{13}-i\zeta d_2N_{23}+i\zeta d_3N_{23})\psi_3=
\zeta^2d^2_2\psi_2,
\end{multline}
\begin{multline}\label{a6}
-\psi_{3,44}+(N_{31,4}+N_{32}N_{21}+i\zeta d_3N_{31}+i\zeta d_1N_{31})\psi_1
+(N_{32,4}+N_{31}N_{12}+i\zeta d_3N_{32}-\\
-i\zeta d_2N_{32})\psi_2++(N_{31}N_{13}+N_{32}N_{23})\psi_3=\zeta^2d^2_3\psi_3.
\end{multline}
Adding the systems (\ref{a1})-(\ref{a3}) and (\ref{a4})-(\ref{a6}),
we obtain in the result
\begin{multline}
-\psi_{1,44}+(N_{12}N_{21}+N_{13}N_{31})\psi_1+(N_{12,4}+N_{13}N_{32})\psi_2+\\
+(N_{13,4}+N_{12}N_{23})\psi_3=\zeta^2d^2_1\psi_1,
\end{multline}
\begin{multline}
-\psi_{2,44}+(N_{21,4}+N_{23}N_{31})\psi_1+(N_{21}N_{12}+N_{23}N_{32})\psi_2+\\
+(N_{23,4}+N_{21}N_{13})\psi_3=\zeta^2d^2_2\psi_2,
\end{multline}
\begin{multline}
-\psi_{3,44}+(N_{31,4}+N_{32}N_{21})\psi_1+(N_{32,4}+N_{31}N_{12}
)\psi_2+\\
+(N_{31}N_{13}+N_{32}N_{23})\psi_3=\zeta^2d^2_3\psi_3.
\end{multline}
It is easy to see that the latter system can be rewritten in the form 
of $3\times 3$ matrix Schr\"{o}dinger equation (\ref{f2}).
\section*{Acknowledgement}
I am deeply grateful to Prof. B. G. Konopelchenko for useful discussions.
  
\end{document}